\documentclass[epj]{webofc}
\usepackage[varg]{txfonts}   
\usepackage{caption}
\usepackage{subcaption}
\wocname{EPJ Web of Conferences}
\woctitle{ICNFP 2016}
\newcommand{\cD}{{\cal D}}

\newcommand{\be}{\begin{equation}}
\newcommand{\ee}{\end{equation}}
\newcommand{\bea}{\begin{eqnarray}}
\newcommand{\eea}{\end{eqnarray}}

\newcommand{\lk}{\left(}

\newcommand{\sli}{\sum\limits}

\newcommand{\il}{\int\limits}

\newcommand{\va}{\vec{a}}
\newcommand{\vB}{\vec{B}}

\newcommand{\vA}{\vec{A}}
\newcommand{\vx}{\vec{x}}
\newcommand{\vy}{\vec{y}}
\newcommand{\vD}{\vec{D}}
\newcommand{\vk}{\vec{k}}

\newcommand{\ii}{\mathrm{i}}
\newcommand{\dd}{\mathrm{d}}
\newcommand{\tr}{\mathrm{tr}}

\renewcommand{\vec}[1]{\mbox{\boldmath$#1$\unboldmath}}

\newcommand{\absvec}[1]{#1}
\begin{document}
\selectlanguage{english}
\title{Hamiltonian approach to QCD in Coulomb gauge: Gribov's confinement scenario at work\footnote{Talk given by H.~Reinhardt at "5th International Conference on New Frontiers in Physics", 6-14 July 2016, Kolymbari, Greece.}}
%
%

\author{H.~Reinhardt\inst{1}\fnsep\thanks{\email{hugo.reinhardt@uni-tuebingen.de}} \and
        G.~Burgio\inst{1}, D.~Campagnari\inst{1}, M.~Quandt\inst{1}, P.~Vastag\inst{1}, 
        H.~Vogt\inst{1} \and E.~Ebadati\inst{1}
}

\institute{Universit\"at T\"ubingen \\
Institut f\"ur Theoretische Physik \\
Auf der Morgenstelle 14 \\
D-72076 T\"ubingen \\
Germany}

\abstract{%
  I will review essential features of the Hamiltonian approach to QCD in Coulomb gauge showing 
  that Gribov's confinement scenario is realized in this gauge. For this purpose I will discuss 
  in detail the emergence of the horizon condition and the Coulomb string tension. I will show 
  that both are induced by center vortex gauge field configurations, which establish the 
  connection between Gribov's confinement scenario and the center vortex picture of confinement. 
  I will then extend the Hamiltonian approach to QCD in Coulomb gauge to finite temperatures, 
  first by the usual grand canonical ensemble and second by the compactification of a spatial 
  dimension. I will present results for the pressure, energy density and interaction measure 
  as well as for the Polyakov loop. 
  }
\maketitle
\section{Introduction}
\label{intro}
V.~N.~Gribov's investigation of the problem of gauge fixing in Yang--Mills theory has led 
to a novel picture of confinement. In this scenario confinement comes about due to the field 
configurations near the Gribov horizon, which give rise to an infrared diverging ghost form factor. 
This approach was worked out in more detail and developed further by D.~Zwanziger and is now 
referred to as the Gribov--Zwanziger confinement mechanism \cite{R3, RZ}. In my talk, I will show 
that this picture is realized in the Hamiltonian approach to Yang--Mills theory in Coulomb gauge. 
Gribov already pointed out that in Coulomb gauge the use of the Hamiltonian approach may be 
advantageous over the usual functional integral approach and this, indeed, turns out to be 
the case. 
The variational approach to the Hamilton formulation of QCD in Coulomb gauge developed in 
refs.~\cite{5,6} has proven to be much more efficient than the formulation of QCD in Coulomb 
gauge by means of the functional integral approach \cite{41}. In my talk I will concentrate 
on a few aspects of the Hamiltonian approach to QCD in Coulomb gauge. I will discuss in detail
the emergence of the horizon condition which is the core of the Gribov--Zwanziger confinement 
scenario. Furthermore, I will show  that the Coulomb string tension is not related to the 
temporal but to the spatial Wilsonian string tension. I will extend then the Hamiltonian 
approach to finite temperatures, first by means of the grand canonical ensemble and second by 
compactification of a spatial dimension.

\section{Hamiltonian approach to Yang--Mills theory in Coulomb gauge at $T = 0$} \label{sec-2}

After canonical quantization and implementation of Gau\ss{}'s law in Weyl and Coulomb gauge, 
the gauge fixed Yang--Mills Hamiltonian reads \cite{3}
\begin{equation}
\label{G1}
H = \frac{1}{2} \int \dd^3 x \lk J^{-1}[A] \vec{\Pi}^a(\vx) \cdot J[A] \vec{\Pi}^a(\vx) +
\vB^a(\vx) \cdot \vB^a(\vx) \right) + H_C \, .
\end{equation}
Here
\begin{equation}
B^a_k(\vx) = \varepsilon_{klm} \left(\partial_l A^a_m(\vx) - \frac{g}{2} f^{abc} A^b_l(\vx)
A^c_m(\vx)\right)
\end{equation}
is the non-Abelian magnetic field, $\Pi^a_k (\vx) = \delta / (\ii \delta A^a_k (\vx))$ 
is the momentum operator, which represents the color electric field and 
\begin{equation}
\label{G2}
J[A] = \mathrm{Det}(-\hat{\vD} \cdot \vec{\partial}) \, , \quad\quad \hat{D}^{a b}_k (\vx) 
= \delta^{a b} \partial^x_k + g \hat{A}_k^{a b}(\vx)
\end{equation}
is the Faddeev--Popov determinant, where $\hat{A}_k^{a b} = f^{a c b} A_k^c$ is the gauge field 
in the adjoint representation. Furthermore,
\begin{equation}
\label{G3}
H_C = \frac{1}{2} \int \dd^3 x \int \dd^3 y \, J[A]^{-1} \rho^a(\vx) J[A] \left[(-\hat{\vD} 
\cdot \vec{\partial})^{-1} (-\vec{\partial}^2) (-\hat{\vD} \cdot 
\vec{\partial})^{-1}\right]^{a b}(\vx, \vy) \rho^b(\vy)
\end{equation}
is the so-called Coulomb term, which arises from the kinetic energy of the longitudinal part 
of the momentum operator after resolving Gau\ss{}'s law. Here,
\begin{equation}
\label{G4}
\rho^a(\vx) = -f^{a b c} \vA^b(\vx) \cdot \vec{\Pi}^c(\vx) + \rho_m^a(\vx)
\end{equation}
is the color charge density,  where the first term on the right-hand side is the contribution 
of the gauge field itself while the last term, $\rho_m$, refers to the matter fields. Due to the 
implementation of Coulomb gauge the scalar product between wave functionals 
\begin{equation}
\label{G5}
\langle \phi | \ldots | \psi \rangle = \int \cD A \, J[A] \phi^*[A] \ldots \psi[A]
\end{equation}
is defined by the functional integral over transversal gauge fields with the 
Faddeev--Popov determinant (\ref{G2}) in the integration measure representing the Jacobian 
of the change of variables from ``Cartesian'' to ``curvilinear'' variables 
in Coulomb gauge. With the gauge fixed Hamiltonian (\ref{G1}) one has to solve the stationary 
Schr\"odinger equation $H \psi[A] = E \psi[A]$ for the vacuum wave functional $\psi[A]$. 
Once $\psi[A]$ is known, all observables and correlation functions can, 
in principle, be calculated. This has been attempted by means of the variational principle 
using Gaussian type ans\"atze for the vacuum wave functional \cite{4,RX}. However, the first 
attempts did not properly include the Faddeev--Popov determinant, which turns out to be  
crucial in order to describe the confinement properties of the theory. Below, I am discussing 
the variational approach developed in refs.~\cite{5, 6}, which differs from previous attempts 
by the ansatz for the vacuum wave functional, the treatment of the Faddeev--Popov determinant 
and, equally important, in the renormalization, see ref.~\cite{Greensite:2011pj} for further 
details.

\subsection{Variational solution of the Schr\"odinger equation}

\begin{figure}
\centering
\includegraphics[width=5cm,clip]{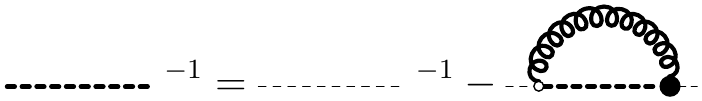}
\caption{Dyson--Schwinger equation for the ghost propagator.}
\label{fig1} %
\end{figure}

\begin{figure}
\centering
\includegraphics[width=7cm,clip]{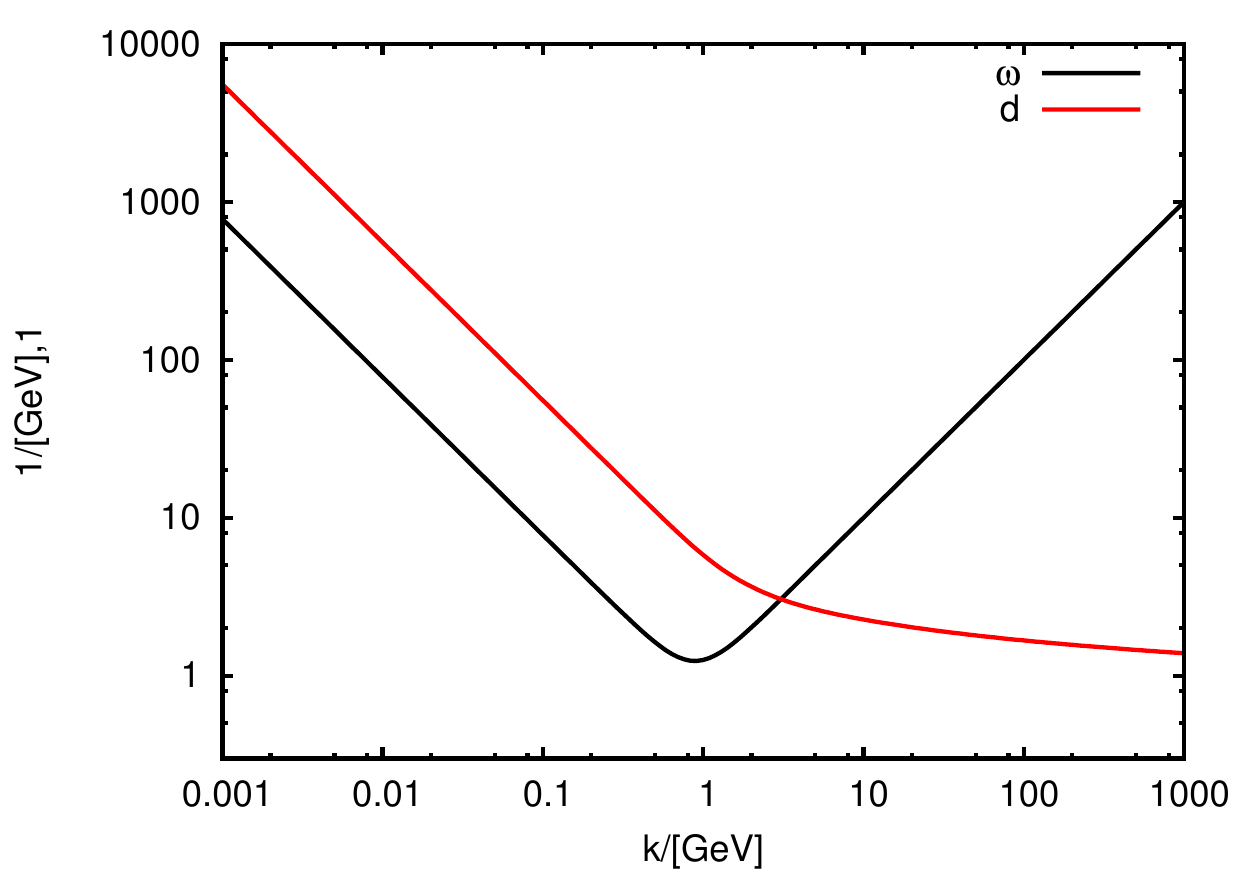}
\caption{Numerical solution of the coupled gap equation for $\omega$ (\ref{G8}) and Dyson--Schwinger 
equation for the ghost form factor $d$ for the renormalization constant $c = 0$ \cite{7}.}
\label{fig2} %
\end{figure}

The ansatz for the vacuum wave functional is inspired by the quantum mechanics of a  particle 
in a spherically symmetric potential for which the ground state wave functional reads
$\psi(r) = u(r) / r$ where the radial wave functional $u(r)$ satisfies a usual one-dimensional 
Schr\"odinger equation and $r$ represents (the square root of the radial part of) the Jacobian 
of the transformation from the Cartesian to spherical coordinates. Our ansatz for the vacuum 
wave functional is given by
\begin{equation}
\label{G6}
\psi[A] = \frac{1}{\sqrt{J[A]}} \exp\left[-\frac{1}{2} \int \dd^3 x \int \dd^3 y \, A_k^a(\vx) 
\omega(\vx, \vy) A_k^a(\vy)\right] \, .
\end{equation}
The inclusion of the pre-exponential factor has the advantage that it eliminates the 
Faddeev--Popov determinant from the integration measure in the scalar product (\ref{G5}). 
Furthermore, for the wave function (\ref{G6}) the gluon propagator is given up to a factor 
of $\frac{1}{2}$ by the inverse of the variational kernel $\omega(\vx, \vy)$. It turns out 
that in the Yang--Mills sector the Coulomb term $H_C$ (\ref{G3}) can be ignored. 
Calculating the expectation value of the remaining parts of the Yang--Mills Hamiltonian 
(\ref{G1}) with the wave functional (\ref{G6}) up to two loops, the minimization of the 
energy density with respect to $\omega(\vx, \vy)$ yields the following gap equation
in momentum space\footnote{Due to translational and rotational invariance, kernels 
such as $\omega(\vx, \vy)$ can be Fourier transformed as 
\[
\omega(\vx, \vy) = \int \frac{d^3 k}{(2\pi)^3}\,e^{i \boldsymbol{k} 
\cdot (\boldsymbol{x}-\boldsymbol{y})}\,\omega(k)\,,
\] 
where the new kernel in momentum space depends on $k = | \vk |$ only. For simplicity, 
we will use the same symbol for the kernel in position and momentum space and go 
back and forth between both representations with impunity.}
\begin{equation}
\label{G8}
\omega^2(k) = \vk^2 + \chi^2(k) + c \, ,
\end{equation}
where $c$ is a finite renormalization constant resulting from the tadpole and
\begin{equation}
\label{G9}
\chi_{k l}^{a b}(\vx, \vy) = -\frac{1}{2} \Bigl\langle \psi \Big\vert \frac{\delta^2 
\ln J[A]}{\delta A_k^a(\vx) \delta A_l^b(\vy)} \Big\vert \psi \Bigr\rangle
\end{equation}
represents the ghost loop. This can be expressed in terms of the ghost propagator
\be
\label{240}
G(\vx, \vy) = \langle \psi \vert {\bigl(-\hat{\vD} \cdot \vec{\partial}\bigr)^{-1}}(\vx, \vy) 
\vert \psi \rangle \, ,
\ee
which is evaluated with the vacuum wave functional (\ref{G6}) in an approximate way 
resulting in a Dyson--Schwinger equation for the form factor
\be
\label{246}
d(\vk) = g \vk^2 G(\vk)
\ee
of the ghost propagator which is diagrammatically illustrated in fig.~\ref{fig1}. This equation 
has to be solved together with the gap equation (\ref{G8}). The numerical solutions are shown 
in fig.~\ref{fig2}.

\subsection{Physical implications of the ghost form factor}

The ghost form factor expresses the deviation of Yang--Mills theory from QED. Dyson--Schwinger
equations are functional differential equations and their solutions are uniquely determined 
after providing appropriate boundary conditions. In the present case the so-called horizon 
condition
\be
\label{272GX}
d^{-1}(0) = 0
\ee
is assumed, which is the key point in Gribov's confinement scenario.

\begin{figure}
\centering
\begin{subfigure}{0.45\textwidth}
\includegraphics[width=\textwidth,clip]{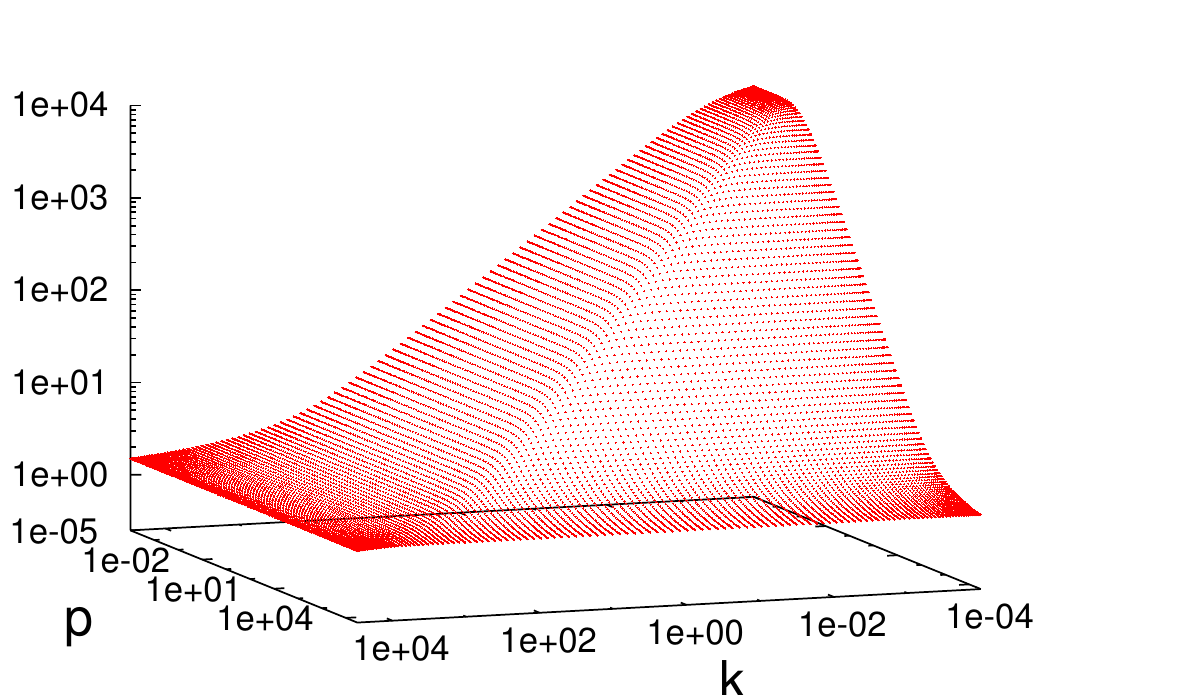}
\caption{}
\end{subfigure}
\quad
\begin{subfigure}{0.45\textwidth}
\includegraphics[width=\textwidth,clip]{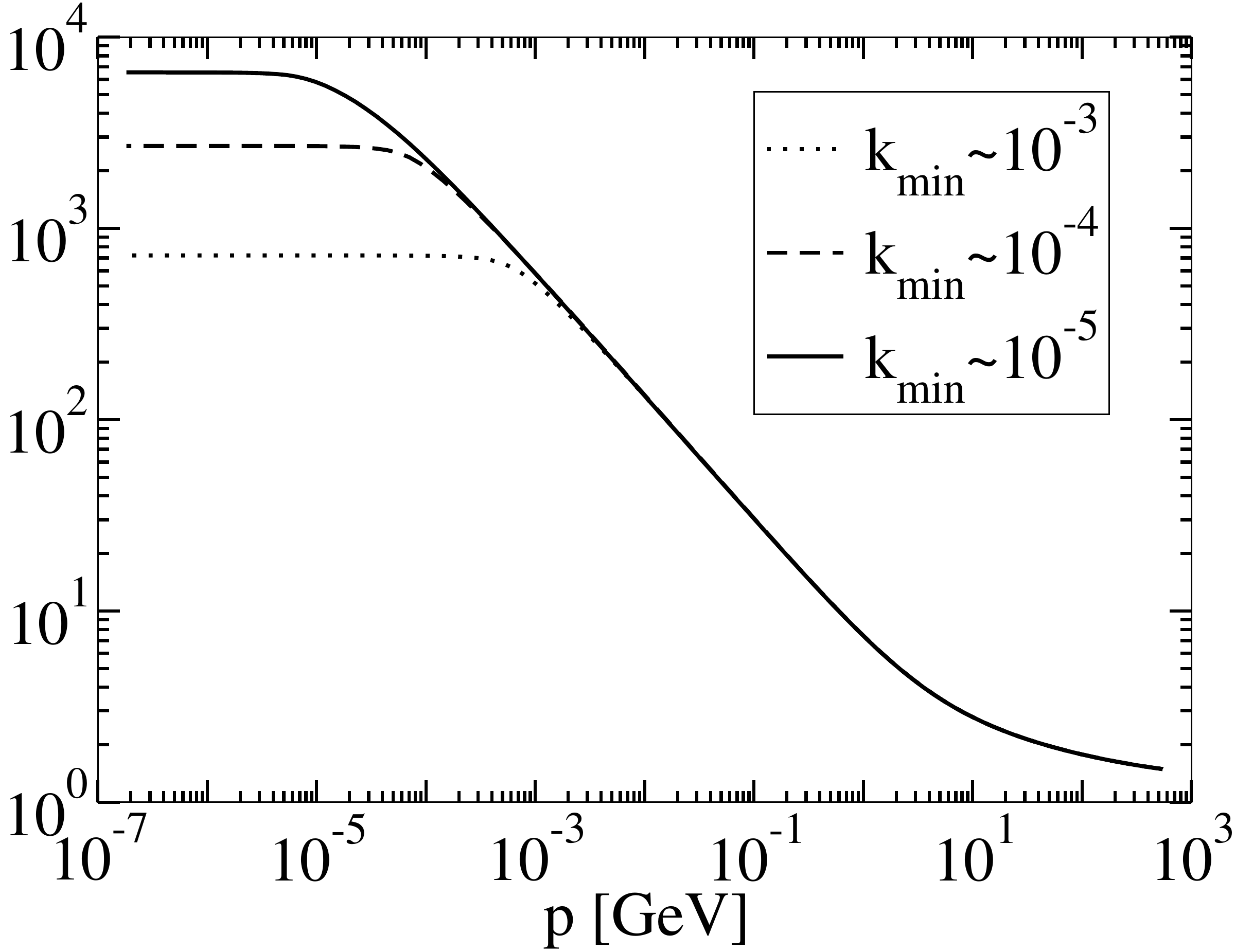}
\caption{}
\end{subfigure}
\caption{(a) The ghost form factor obtained in ref.~\cite{Leder:2010ji} from the solution 
of the renormalization group flow equations. Here $p$ represents the momentum variable of 
the ghost form factor while $k$ is the infrared momentum cutoff of the flow equations. 
(b) Cuts through subfigure (a) at various values of the momentum scale $k$ of the flow equations.}
\label{fig3}%
\end{figure}

Alternatively to the variational approach, one can indirectly determine the vacuum wave 
functional by solving the functional renormalization group flow equations for the various 
propagators and functions of the Hamiltonian approach. Restricting the flow equations to 
those for the ghost and gluon propagators, one finds for the ghost form factor the result 
shown in fig.~\ref{fig3}. Starting with a constant ghost form factor in the ultraviolet, 
the ghost form factor develops an infrared singularity as the momentum cutoff of the flow 
equation tends to zero. This is nicely seen in fig.~\ref{fig3} (b), which shows a cut 
through fig.~\ref{fig3} (a) at fixed renormalization group scale $k$.

\begin{figure}
\centering
\includegraphics[width=7cm,clip]{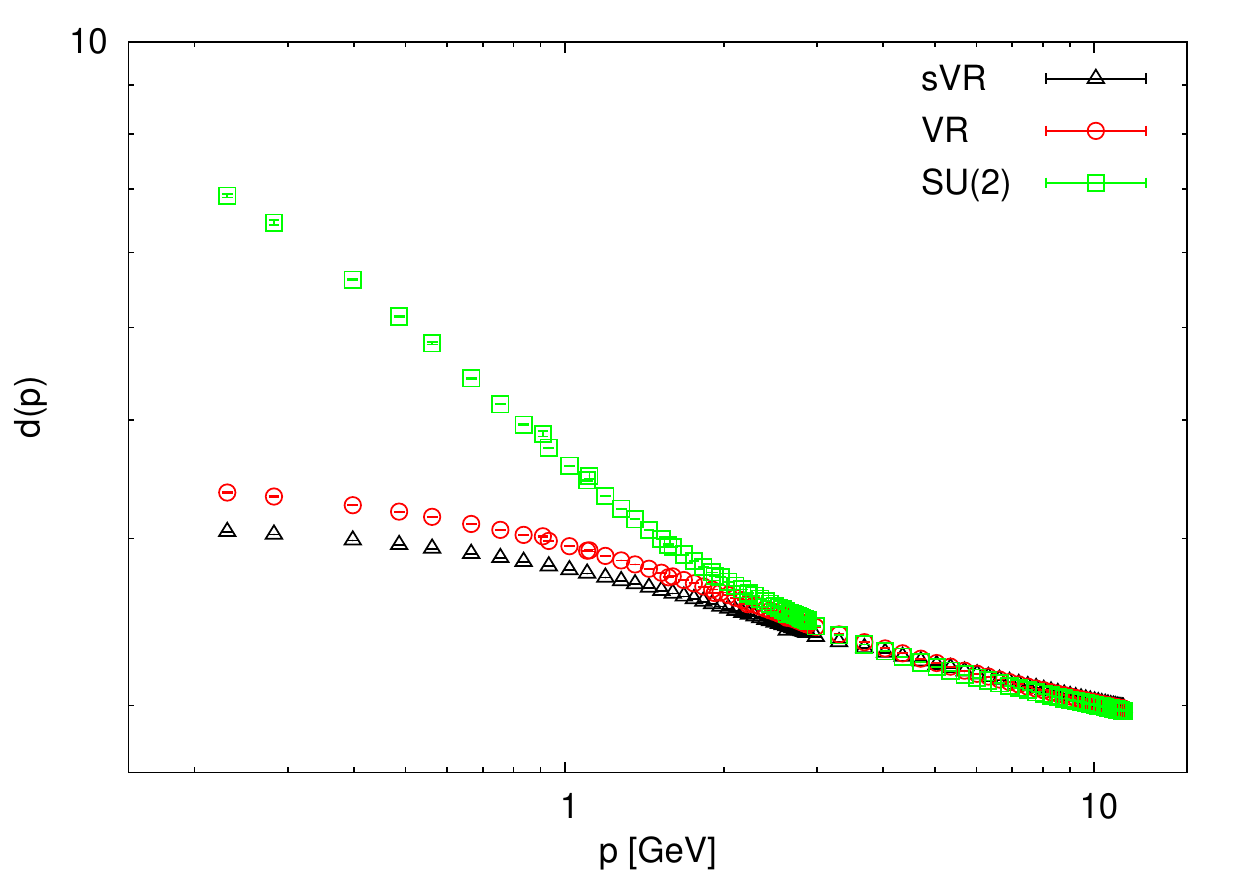}
\caption{The ghost form factor in Coulomb gauge calculated on the lattice in ref.~\cite{Burgio:2015hsa}  
(green symbols). The red and black symbols show the results obtained for the ghost form factor 
when all center vortices or the spatial center vortices are removed from the ensemble of gauge 
field configurations, see text.}
\label{fig4} %
\end{figure}

Let us also mention that the horizon condition (\ref{272GX}) need not be assumed in the 
case of $D = 2 + 1$ dimensions, where it is a necessary consequence of the coupled equations 
for the ghost and gluon propagators obtained with the variational principle. Finally, 
the horizon condition (\ref{272GX}) is also seen in the lattice data for the ghost form factor 
shown in fig.~\ref{fig4}.

Coulomb gauge is called a physical gauge since in QED the remaining transversal components 
are the gauge invariant degrees of freedom. This is not the case for Yang--Mills theory. 
However, Coulomb gauge can be viewed as a physical gauge also in the case of Yang--Mills theory
in the sense that the inverse ghost form factor in Coulomb gauge represents the 
dielectric function of the Yang--Mills vacuum \cite{9}
\begin{equation}
\label{G12}
\epsilon(k) = d^{-1}(k) \, .
\end{equation}
The horizon condition (\ref{272GX}) guarantees that this function vanishes in the infrared, 
$\epsilon(k = 0) = 0$. This implies that the Yang--Mills vacuum is a perfect color dielectricum, 
i.e.~a dual superconductor. In this way the Hamiltonian approach in Coulomb gauge relates Gribov's 
confinement scenario to the dual Mei{\ss}ner effect, a confinement mechanism realized through 
the condensation of magnetic monopoles and proposed by Mandelstam and 't Hooft \cite{10,11}. 
The dielectric function obtained here as inverse ghost form factor is also in accord with the 
phenomenological bag model picture of hadrons. Inside the hadron, i.e.~at small distance, 
the dielectric function is that of a normal vacuum while outside the physical hadrons the 
vanishing of the dielectric constant implies the absence of free charges by Gau\ss{}'s law.

One may now ask, what field configurations induce the horizon condition and thus confinement? 
Given the relation of Gribov's confinement scenario to the dual superconductor, we expect magnetic 
monopoles to play a substantial role. Lattice calculations carried out in the so-called indirect 
maximum center gauge, which contains the maximum Abelian gauge in an intermediate step, show that 
magnetic monopoles are tied to center vortices. The latter are string-like gauge field configurations 
for which the Wilson loop equals a non-trivial center element of the gauge group, provided the 
loop has non-trivially linking with the center vortex string. When the center vortex content 
of the gauge field configurations is removed, which can be done on the lattice, one finds that the 
Wilsonian string tension and thus confinement disappears \cite{deForcrand:1999our}. Figure \ref{fig4} 
shows the ghost loop obtained on the lattice when the center vortices are removed from the ensemble 
of gauge field configurations \cite{Burgio:2015hsa}. The ghost loop becomes infrared flat and the
horizon condition is lost. This shows that center vortices induce the horizon condition which is 
the corner stone of Gribov's confinement scenario. This also shows that Gribov's confinement 
scenario is tied to the center vortex picture of confinement. This is in accord with the 
observation that center vortices and magnetic monopoles are located on the Gribov horizon 
of Coulomb gauge \cite{Greensite:2004ke}.

\subsection{Comparison with lattice calculation}

\begin{figure}
\centering
\begin{subfigure}{0.45\textwidth}
\includegraphics[width=\textwidth,clip]{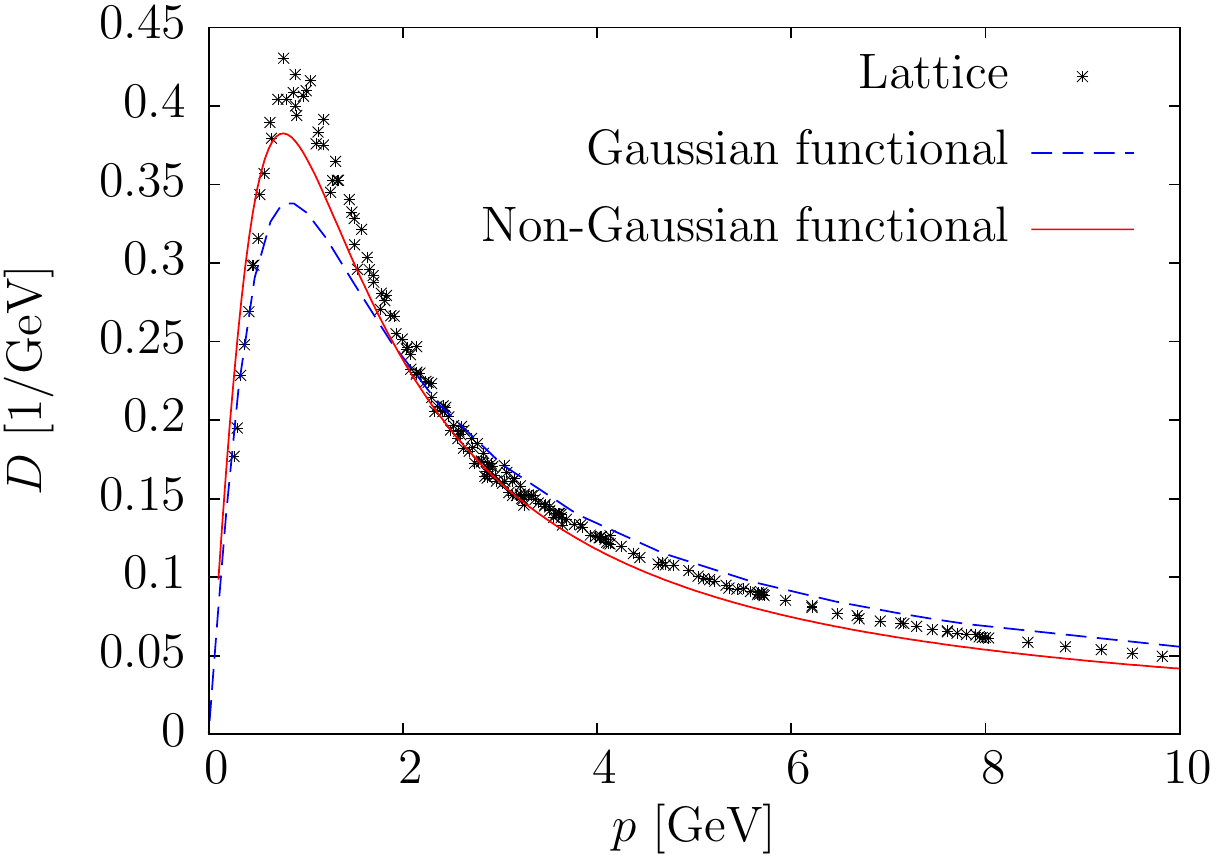}
\caption{}
\end{subfigure}
\quad
\begin{subfigure}{0.45\textwidth}
\includegraphics[width=\textwidth,clip]{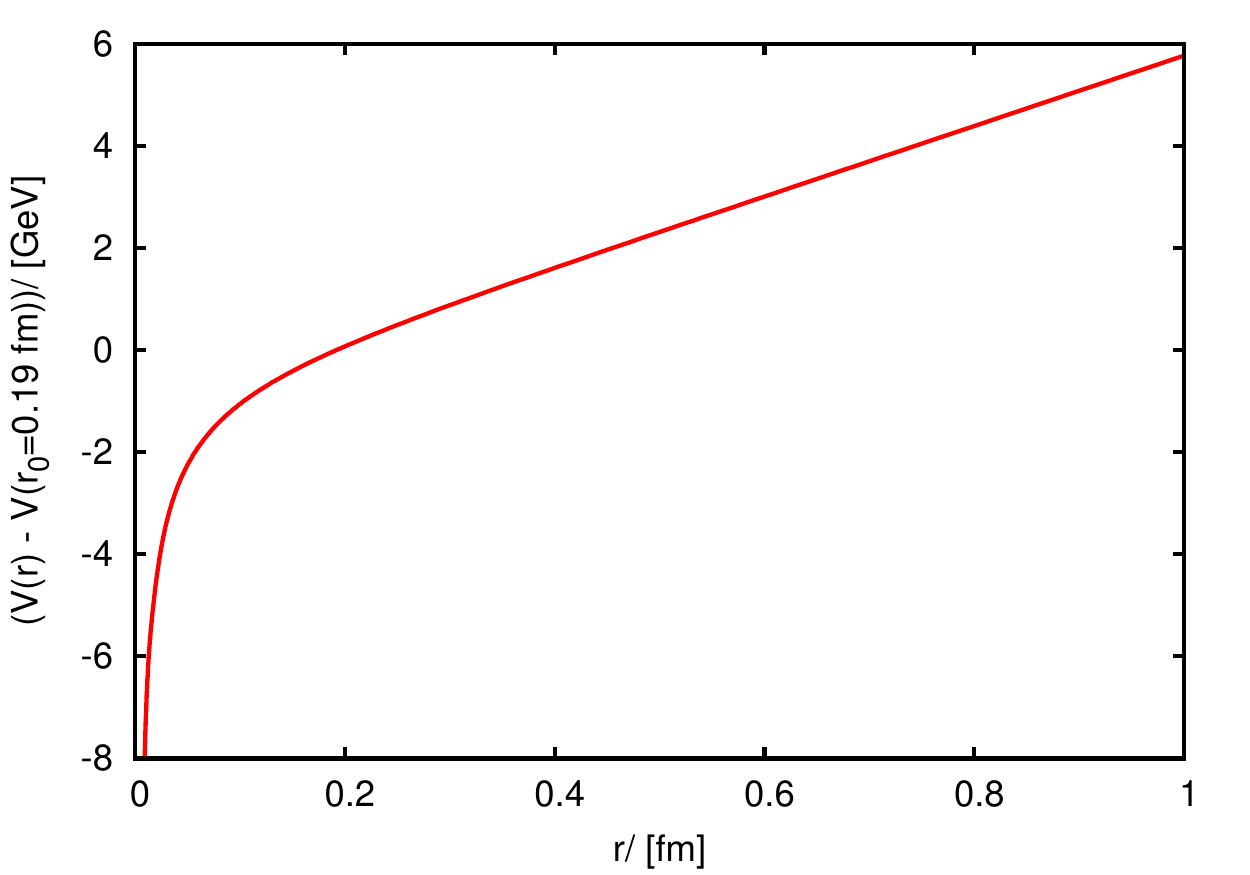}
\caption{}
\end{subfigure}
\caption{(a) The static gluon propagator in Coulomb gauge calculated on the lattice for SU(2) 
gauge theory (crosses). The dashed and the full curves show the result of the variational 
calculation using, respectively, a Gaussian and non-Gaussian ansatz for the vacuum wave functional. 
(b) The non-Abelian Coulomb potential (\ref{G15}) obtained in the variational approach \cite{7}.}
\label{fig5}%
\end{figure}

Let us now compare the results of the variational solution with lattice calculations \cite{13}. 
Figure \ref{fig5} (a) shows the gluon propagator of Coulomb gauge obtained in SU(2) gauge theory.
It is remarkable that the lattice data can be nicely fitted by Gribov's formula \cite{R3}
\begin{equation}
\label{G13}
\omega(p) = \sqrt{p^2 + \frac{M^4}{p^2}}
\end{equation}
where $M$ is the so-called Gribov mass. The variational calculations reproduce the infrared behavior 
of the lattice propagator perfectly and are also in reasonably agreement with the lattice data 
in the ultraviolet. However, in the mid-momentum regime some strength is missing in the variational 
calculation. This missing strength is the result of the Gaussian type ansatz for the vacuum 
wave functional. In ref.~\cite{14}, the ansatz for the vacuum wave functional was extended to 
include also cubic and quartic terms of the gauge field in the exponent of the vacuum wave functional
\begin{subequations}
\begin{align}
\psi[A] &\sim \exp\bigl[-S[A]\bigr] \, , \\
S[A] &= \frac{1}{2} \int A \omega A + \frac{1}{3!} \int \gamma^{(3)} A A A + \frac{1}{4!} 
\int \gamma^{(4)} A A A A
\end{align}
\end{subequations}
and one finds the full  curve in fig.~\ref{fig5} (a), which gives a much better agreement
with the lattice data in the mid-momentum regime.

\subsection{The Coulomb string tension}

\begin{figure}
\centering
\begin{subfigure}{0.45\textwidth}
\includegraphics[width=\textwidth,clip]{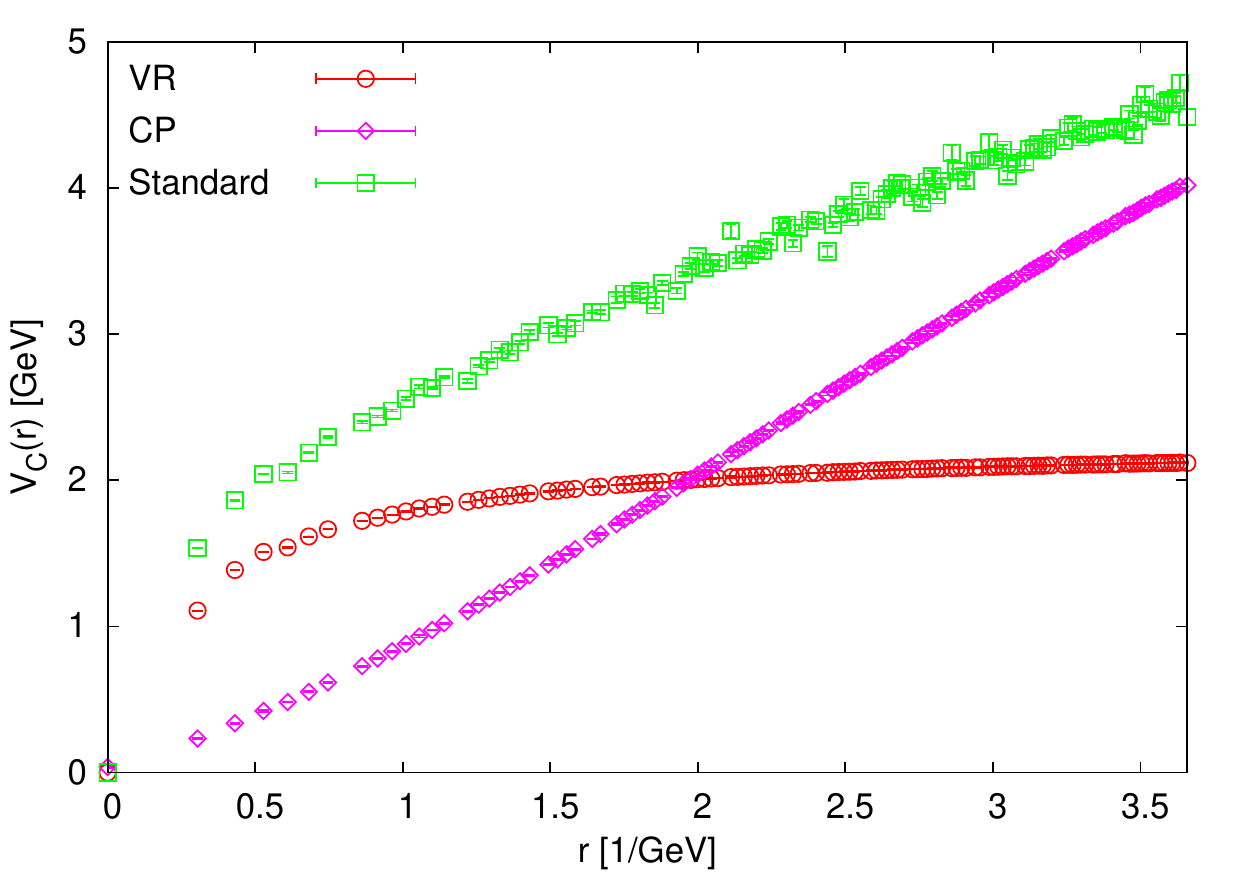}
\caption{}
\end{subfigure}
\quad
\begin{subfigure}{0.45\textwidth}
\includegraphics[width=\textwidth,clip]{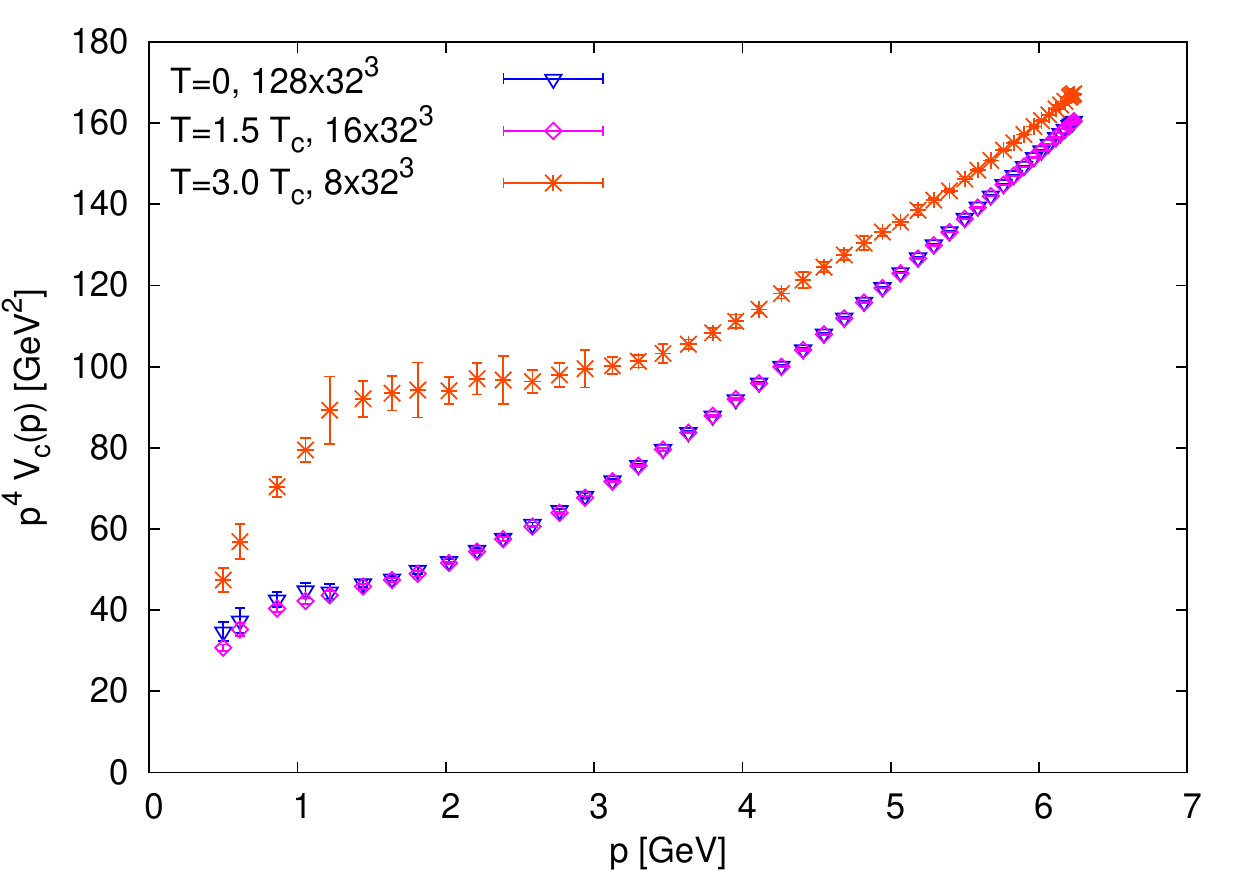}
\caption{}
\end{subfigure}
\caption{(a) Non-Abelian Coulomb potential $V_C$ in coordinate space calculated on the 
lattice (green symbols). Red and violet symbols show the results obtained by vortex removal 
and center projection, respectively. (b) Measurements for the quantity $p^4 V_C(p)$ as 
function of the momentum $p$ for various temperatures \cite{Burgio:2015hsa}.}
\label{fig8}%
\end{figure}

So far we have ignored the Coulomb Hamiltonian $H_C$ (\ref{G3}) which is reasonable in the 
Yang--Mills sector. However, this term becomes extremely important in the quark sector. 
Let us consider the piece of $H_C$ quadratic in the color density of the matter fields $\rho_m$. 
From this term the Faddeev--Popov determinant drops out and its Yang--Mills vacuum expectation 
value provides the static potential for the matter fields
\begin{equation}
\label{G15}
V_C(|\vx - \vy|) = g^2 \Bigl\langle \psi \Bigr\vert \left[\bigl(-\hat{\vD} \cdot 
\vec{\partial}\bigr)^{- 1} \bigl(-\vec{\partial}^2\bigr) \bigl(-\hat{\vD} \cdot 
\vec{\partial}\bigr)^{-1}\right](\vx, \vy) \Bigl\vert \psi \Bigr\rangle \, ,
\end{equation}
which is referred to as the non-Abelian Coulomb potential. The Coulomb potential resulting from
the variational approach is shown in fig.~\ref{fig5} (b). At small distance it behaves like 
an ordinary Coulomb potential but rises linearly at large distance with a coefficient given
by the so-called Coulomb string tension $\sigma_C$, which can be shown to represent an upper
bound to the Wilsonian string tension $\sigma_W$ \cite{R2}. On the lattice one finds a value of 
$\sigma_C = 2 \ldots 4 \, \sigma_W$ \cite{Greensite:2004ke, Burgio:2015hsa, R4a}. 
At finite temperatures the Wilsonian string tension $\sigma_W$ splits into a spatial and a 
temporal one. Above the deconfinement phase transition, the temporal 
string tension $\sigma_t$ vanishes while the spatial string tension $\sigma_s$ even 
slightly increases with the temperature. On the lattice one finds that the temporal and spatial string 
tension are induced by the spatial and temporal, respectively, center vortex content of the 
gauge field. When one removes the spatial center vortices from the ensemble of gauge fields, 
the temporal string tension is untouched while the spatial string tension disappears. 
Figure \ref{fig8} (a) shows the non-Abelian Coulomb potential calculated on the lattice. 
When one removes the spatial center vortices this Coulomb potential becomes infrared flat, 
i.e.~the Coulomb string tension $\sigma_C$ disappears. This shows that the Coulomb string 
tension is tied to the spatial (Wilsonian) string tension and not to the temporal one, because both 
are sensitive to the same underlying degrees of freedom (the spatial center vortices). 
Since the spatial string tension increases with the temperature above the critical temperature
we should expect the same behavior for the Coulomb string tension. This, indeed, is 
observed on the lattice as can be seen from fig.~\ref{fig8} (b), where the quantity 
$p^4 V_C(p)$ is shown for various temperatures. As can be seen from this figure, the 
Coulomb string tension
\begin{equation}
\sigma_C = \lim\limits_{p \to 0} p^4 V_C(p)
\end{equation}
remains more or less constant as the temperature is increased from $0$ to $1.5$ 
times the critical temperature, and then increases strongly when further increasing the  
temperature to 3 times the critical one. This is in accordance with the temperature 
behavior of the spatial string tension found in lattice calculations \cite{x6}.


The variational approach to Yang--Mills theory developed in refs.~\cite{5,6} has also been 
extended to full QCD, see refs.~\cite{23,24,19}. Due to lack of time I will not consider the
quark sector within my talk.

\section{Hamiltonian approach to Yang--Mills theory at finite temperatures}

\subsection{Grand canonical ensemble} \label{subsec-2}

\begin{figure}
\centering
\begin{subfigure}{0.45\textwidth}
\includegraphics[width=\textwidth,clip]{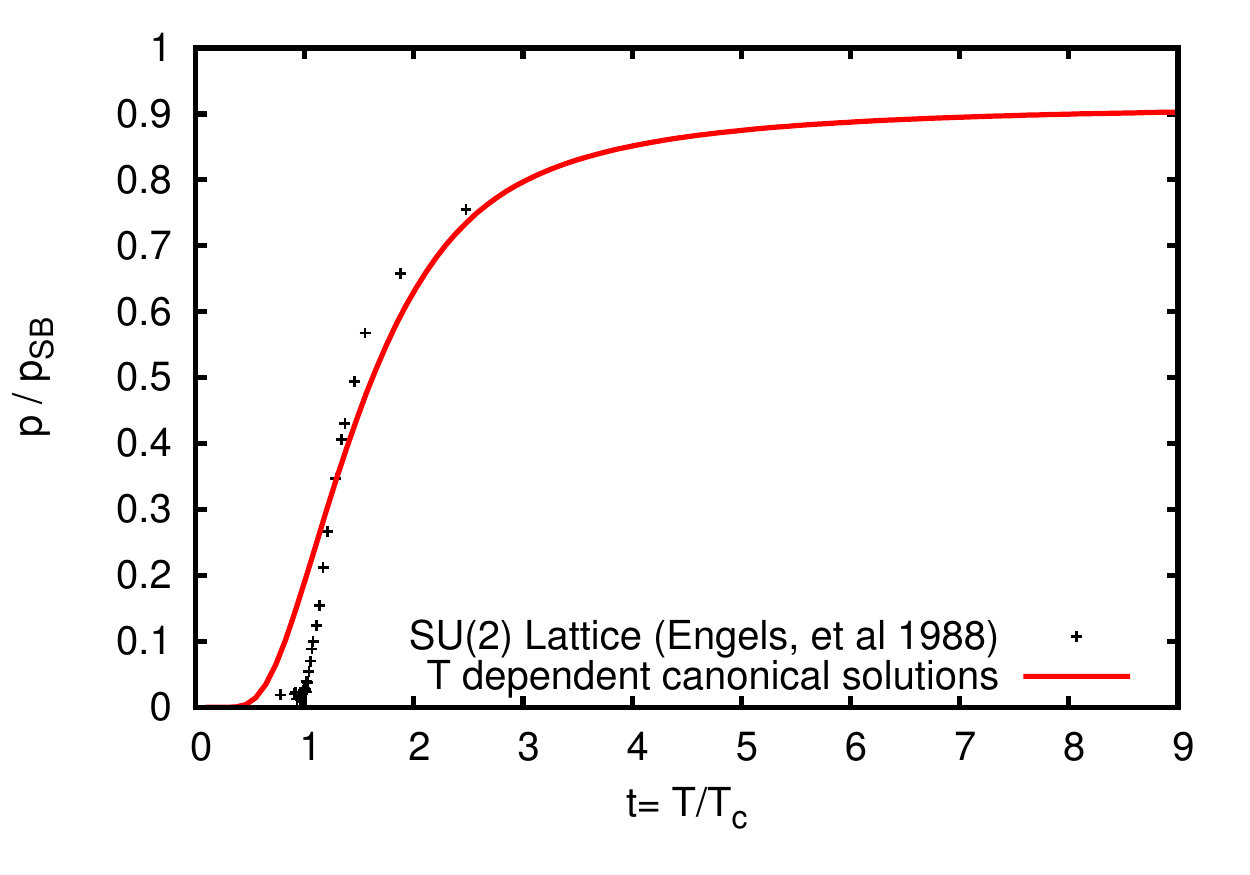}
\caption{}
\end{subfigure}
\quad
\begin{subfigure}{0.45\textwidth}
\includegraphics[width=\textwidth,clip]{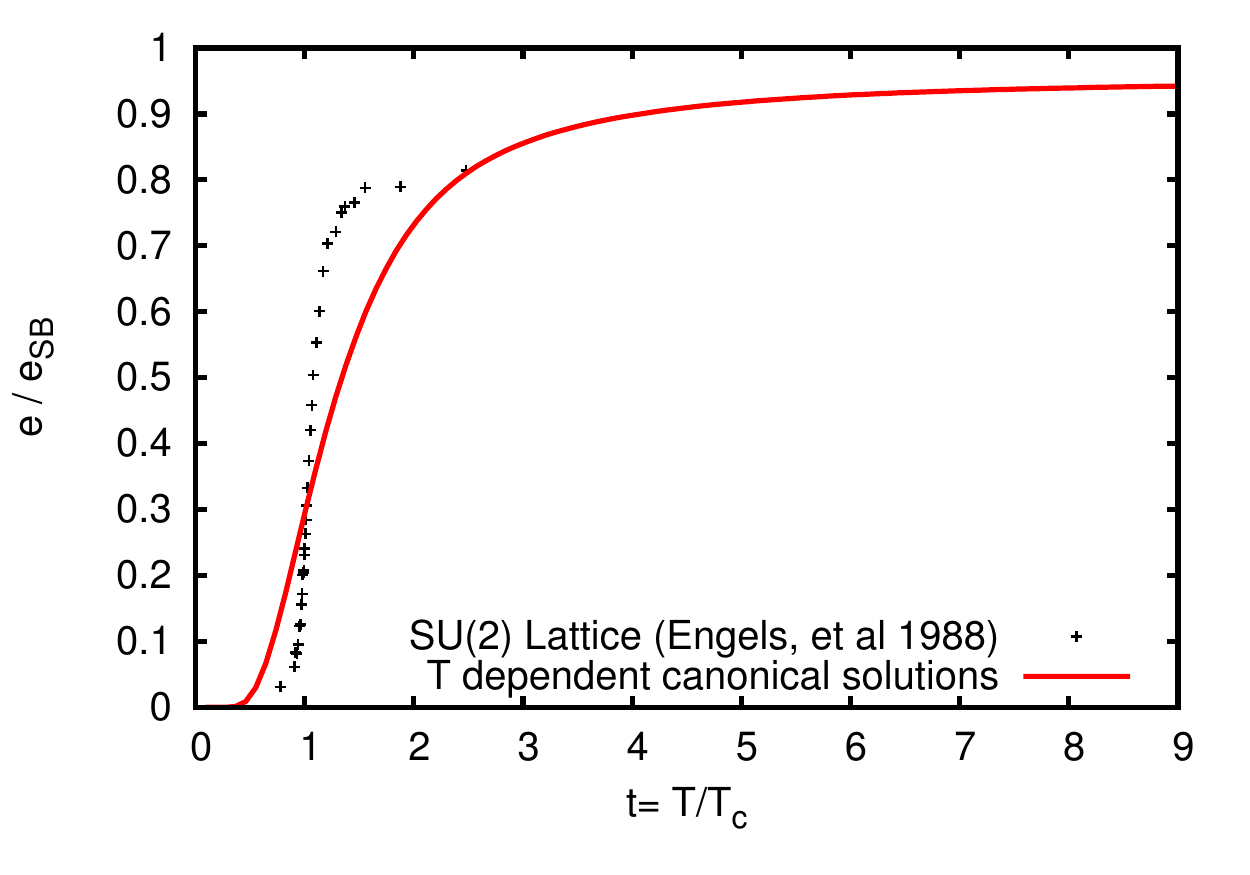}
\caption{}
\end{subfigure} \\
\begin{subfigure}{0.45\textwidth}
\includegraphics[width=\textwidth,clip]{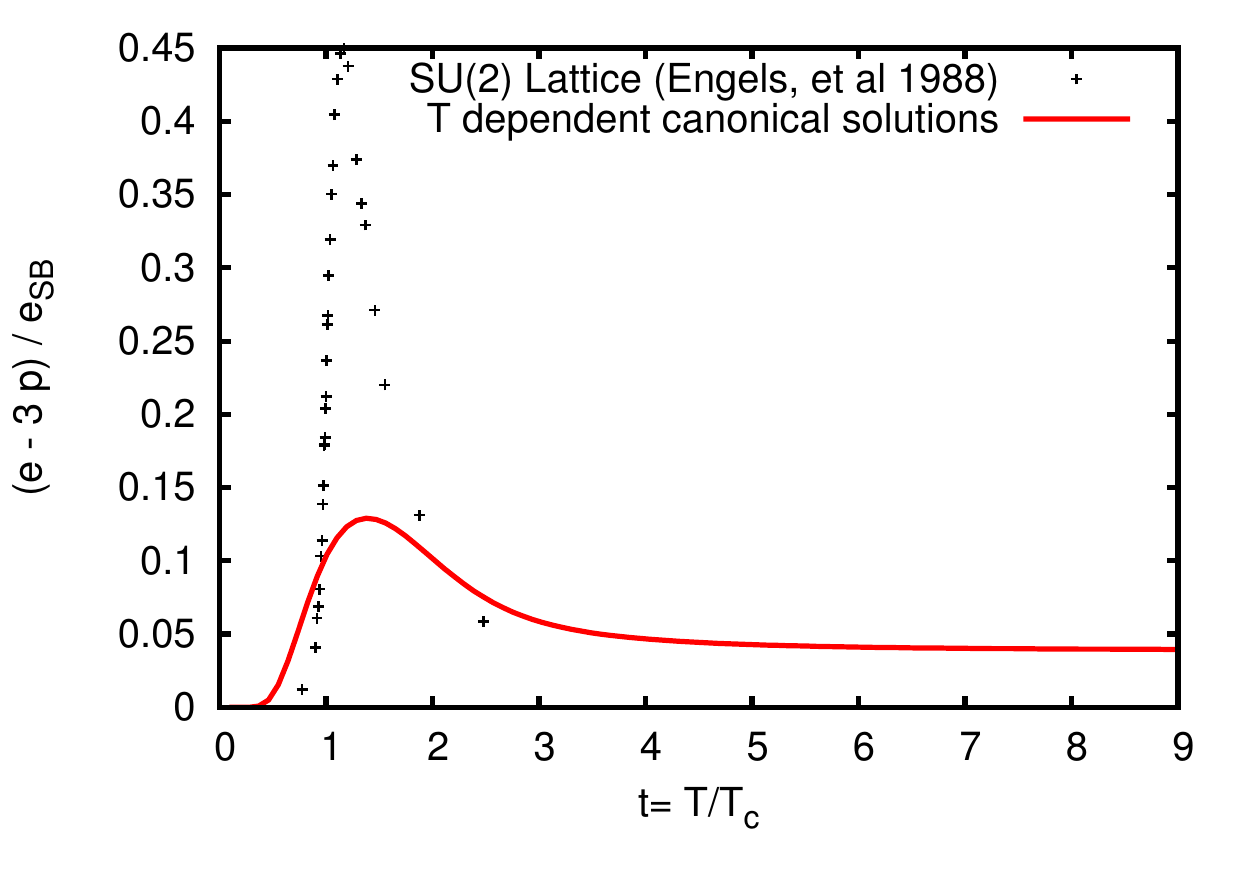}
\caption{}
\end{subfigure}
\caption{The (a) pressure, (b) energy density and (c) interaction strength of SU$(2)$ 
gluon dynamics calculated in the finite temperature variational approach to Yang--Mills 
theory in Coulomb gauge developed in refs.~\cite{26,8}.}
\label{fig10}%
\end{figure}

In refs.~\cite{26,8}, the variational approach to Yang--Mills theory in Coulomb gauge 
was extended to finite temperatures $T$ by making a quasi-particle ansatz for the 
density matrix of the grand canonical ensemble with zero chemical potential
\begin{equation}
\label{G16}
\tilde{D} = \exp\left(-\tilde{H} / T\right) \, ,
\end{equation}
where $\tilde{H}$ denotes the quasi-gluon (single-particle) Hamiltonian whose 
quasi-particle energies were determined by minimizing the free energy
\begin{equation}
\label{G17}
F = \langle H \rangle_T - T S \to \min.
\end{equation}
Here,
\begin{equation}
\label{G18}
\langle \ldots \rangle_T = \frac{\tr \left(\tilde{D} \ldots\right)}{\tr \, \tilde{D}}
\end{equation}
denotes the thermal expectation value of the grand canonical ensemble (i.e.~the trace 
extends over the whole gluonic Fock space) and
\begin{equation}
\label{G19}
S = -\tr\left(\frac{\tilde{D}}{\tr \, \tilde{D}} \ln \frac{\tilde{D}}{\tr \, 
\tilde{D}}\right) = - \Bigg\langle \ln \frac{\tilde{D}}{\tr \, \tilde{D}} \Bigg\rangle_T
\end{equation}
is the usual entropy defined with the density matrix (\ref{G16}). The variational 
principle (\ref{G17}) has been solved in ref.~\cite{8} analogously to the one at
zero temperature. Recently this approach has been used to calculate the pressure, 
energy density and interaction strength of SU(2) gluon dynamics \cite{47}. 
The results are shown in fig.~\ref{fig10}.
The agreement with lattice data is not very impressive but one should keep in mind that 
these are microscopic self-consistent calculations.

In ref.~\cite{Zwanziger:2004np}, gluon dynamics has been studied in a quasi-particle
picture using Gribov's formula (\ref{G13}) for the gluon quasi-particle energy. 
Thereby the Gribov mass was adjusted to reproduce the high temperature tail of the 
interaction strength yielding a Gribov mass of $M = 705 \, \mathrm{MeV}$ which is 
substantially smaller than the value $M = 880 \, \mathrm{MeV}$ found on the lattice \cite{13}.
It is therefore not surprising that the model calculation of ref.~\cite{Zwanziger:2004np} 
did not properly describe the interaction strength in the low temperature region. 
If one wants to reproduce pressure, energy density or interaction strength of 
Yang--Mills theory within the quasi-particle model used in ref.~\cite{Zwanziger:2004np}, 
a temperature dependent Gribov mass is required. This is seen from the lattice 
calculation of the gluon propagator $D(p)$ at finite temperature presented in 
fig.~\ref{fig11}, where the quantity $D(p) / p$ is shown. Assuming Gribov's 
formula (\ref{G13}) for the gluon energy at all temperatures, this quantity 
reduces in the zero momentum limit to $1 / (2 M^2)$. As can be seen from 
fig.~\ref{fig11}, the Gribov mass $M$ does not change much as one increases the 
temperature from $0$ to $1.5 T_C$ but increases substantially when the temperature 
is further increased to 3 times the critical one.

\subsection{Finite temperatures by compactification of a spatial dimension}

There is a more efficient way to treat Yang--Mills theory at finite temperature 
within the Hamiltonian approach. The motivation comes from the Polyakov loop
\begin{equation}
\label{G20}
P[A_0](\vx) = \frac{1}{d_r} \tr P \exp\left[\ii \il_0^L \dd x^4 \, A_0 (\vx, x^4) \right] \, ,
\end{equation}
where $d_r$ denotes the dimension of the representation of the gauge group and $L$ 
is the inverse temperature. \footnote{Recall that in the Euclidean functional integral 
approach, finite temperatures are introduced by compactifying the time axis.} The expectation 
value of the Polyakov loop $\langle P[A_0] \rangle$ is an order parameter for 
confinement \cite{R5}. In the center symmetric confined phase this quantity vanishes
while it approaches unity in the deconfined phase. This quantity cannot be calculated 
straightforwardly in the Hamiltonian approach due to the unrestricted time interval and 
the use of the Weyl gauge $A_0 = 0$. One can, however, exploit $O(4)$ invariance and 
interchange the Euclidean time axis 
with a spatial axis. The temporal (anti-)periodic boundary conditions to the fields 
become then spatial boundary conditions while the new (Euclidean) time axis has infinite 
extent, as is required for the Hamiltonian approach. 
As the result, the partition function is entirely given by the ground state 
calculated on the spatial manifold $\mathbb{R}^2 \times S^1(L)$ where $L$ is again 
the inverse temperature. The whole thermodynamics is then encoded in the vacuum 
calculated on the partially compactified spatial manifold $\mathbb{R}^2 \times S^1(L)$. 
This approach has been worked out in ref.~\cite{Reinhardt:2016xci} and was used in 
refs.~\cite{27,28} to calculate the Polyakov loop within the Hamiltonian approach.

In refs.~\cite{31,32} it was shown that instead of using the expectation value of 
the Polyakov loop as order parameter of confinement one can also use the Polyakov
loop of the expectation value $P[\langle A_0 \rangle]$ or the expectation value of 
the gauge field $\langle A_0(\vx) \rangle$ whereas the gauge field has to be in 
the Polyakov gauge, i.e.~it has to be independent of the (Euclidean) time and 
diagonal in color space. Using this result it turns out that the most efficient 
way to obtain the Polyakov loop is to calculate the effective potential of a 
temporal background field $a_0$ fulfilling the Polyakov gauge and calculate the 
Polyakov loop from the minimum of this effective potential, say $\bar{a}_0$. The 
quantity $P[\bar{a}_0]$ can then be used as alternative to $\langle P [A_0] \rangle$. 
This was done within the Hamiltonian approach in refs.~\cite{27,28} using the 
finite temperature formulation of quantum field theory developed in 
ref.~\cite{Reinhardt:2016xci}. The color diagonal constant background field 
$\va = \va_3 t_3$ ($t$ denotes the generator of the color group in the fundamental 
representation) has to be directed along the compactified spatial axis. Using 
the zero temperature gluon and ghost propagator as input one finds within this 
approach for SU(2) the effective potential $e [a]$ shown in fig.~\ref{fig12} (a).
The potential shows a second order phase transition. Using the Gribov mass of 
$M = 880 \, \mathrm{MeV}$ for fixing the scale one finds from the self-consistent 
(zero temperature) solution a critical temperature of $T_C = 269 \, \mathrm{MeV}$.
\begin{figure}
\centering
\scalebox{.68}{\input{fig11}}
\caption{$D(p) / p$ for $T = 0,\, 1.5 T_C\text{ and } 3 T_C$ \cite{42}.}
\label{fig11} %
\end{figure} %

\begin{figure}
\centering
\begin{subfigure}{0.45\textwidth}
\includegraphics[width=\textwidth,clip]{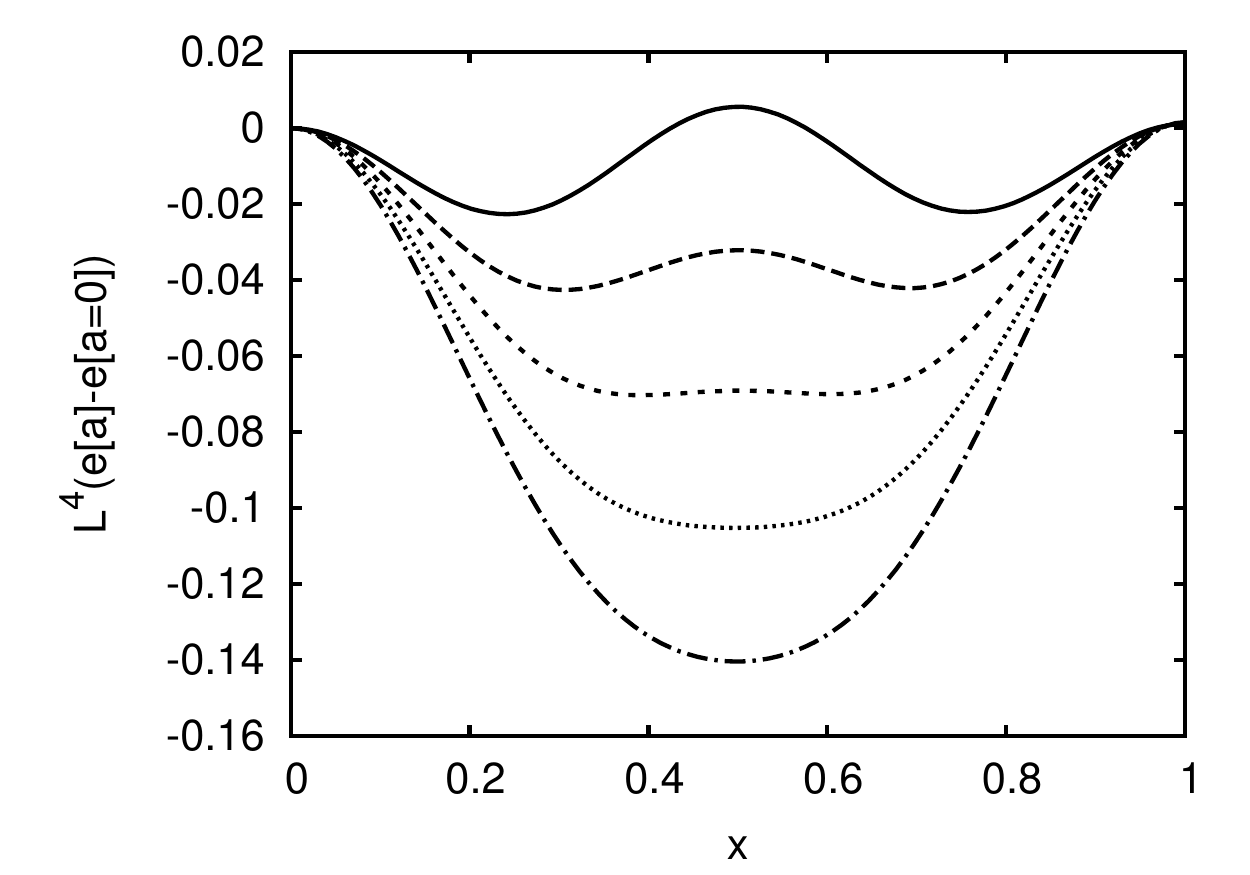}
\caption{}
\end{subfigure}
\quad
\begin{subfigure}{0.45\textwidth}
\includegraphics[width=\textwidth,clip]{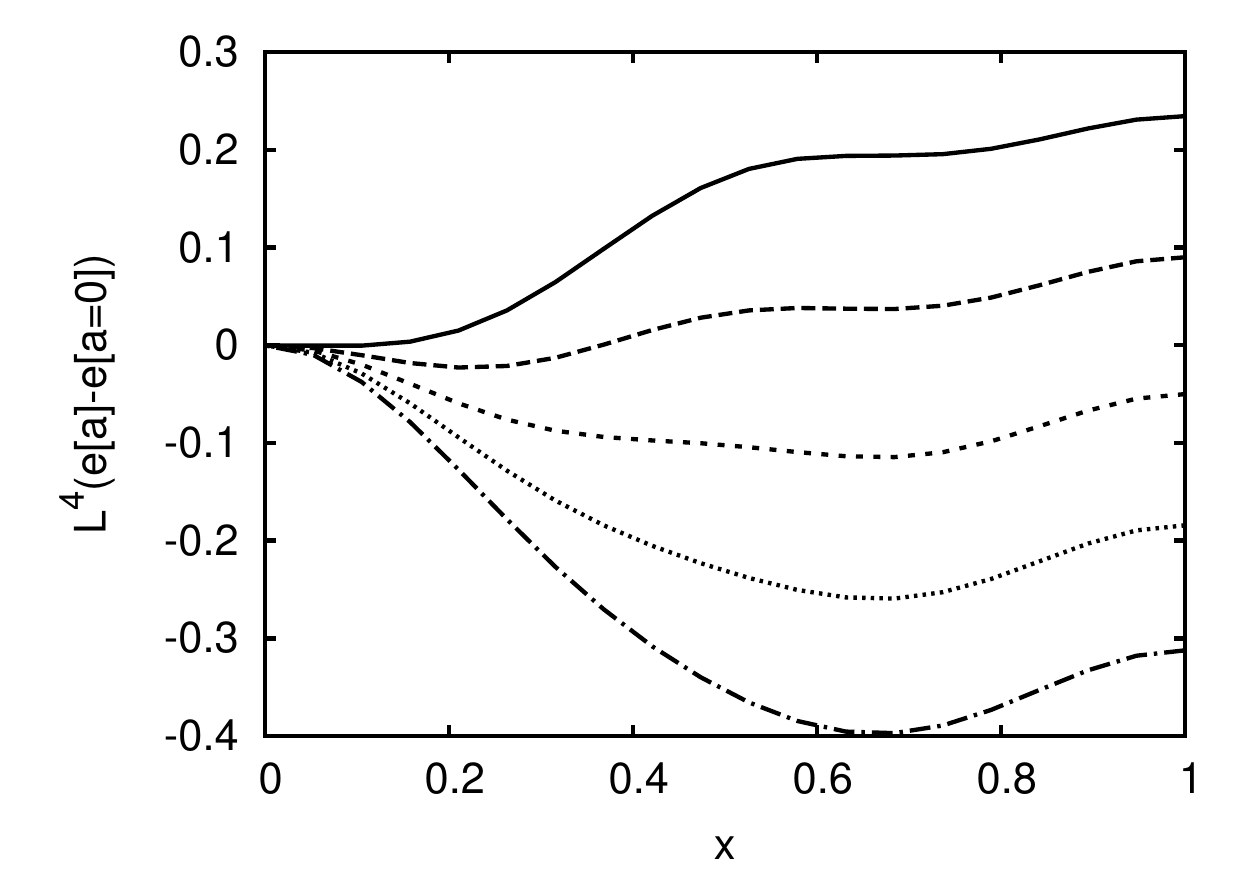}
\caption{}
\end{subfigure}
\caption{(a) The effective potential of the Polyakov loop (more precisely of the background 
field along the compactified spatial dimension) for the case of SU(2) for various temperatures 
(increasing from bottom to top) near the deconfinement phase transition; $x = \frac{a_3 L}{2 \pi}$. 
(b) The $a_8 = 0$ section of the effective potential of the Polyakov loop for the gauge group 
SU(3) for various temperatures increasing from bottom to top, see also fig.~\ref{fig13}.}
\label{fig12}%
\end{figure}%

\begin{figure}
\centering
\begin{subfigure}{0.45\textwidth}
\includegraphics[width=\textwidth,clip]{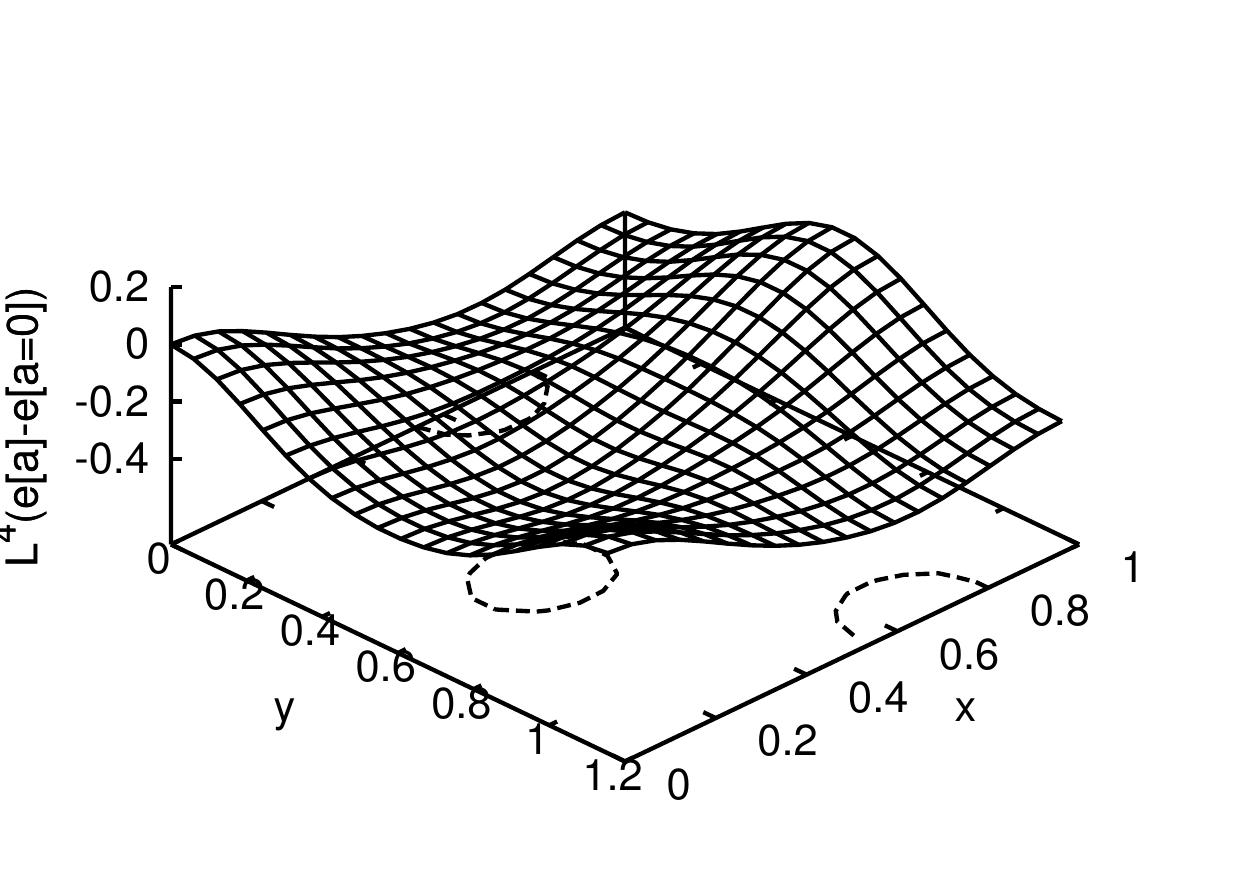}
\caption{}
\end{subfigure}
\quad
\begin{subfigure}{0.45\textwidth}
\includegraphics[width=\textwidth,clip]{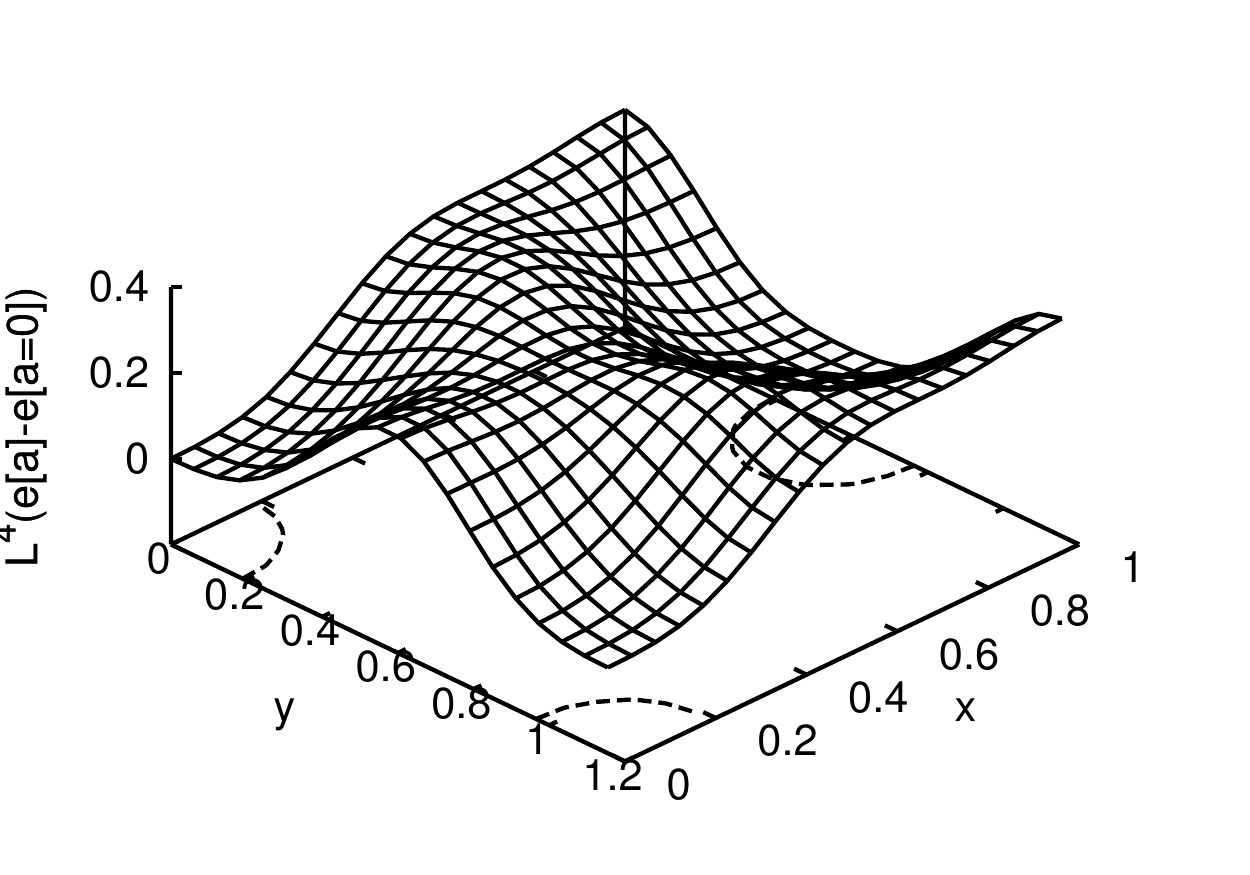}
\caption{}
\end{subfigure}
\caption{The effective potential of the Polyakov loop as function of the gauge fields 
along the Cartan generators $a_3$ and $a_8$ (a) for a temperature below the critical 
one and (b) for a temperature above the critical one;
$x = \frac{a_3 L}{2 \pi}$, $y = \frac{a_8 L}{2 \pi}$.}
\label{fig13}%
\end{figure}

Since the SU(3) algebra consists of three SU(2) algebrae characterized by the three 
non-zero positive roots
\begin{equation}
\label{G21}
\vec{\sigma} = \bigl(1, 0\bigr) \, , \quad \left( \frac{1}{2}, \frac{1}{2}
\sqrt{3} \right) \, , \quad  \quad \left( \frac{1}{2}, - \frac{1}{2} \sqrt{3} \right)
\end{equation}
the effective potential for SU(3) is given by the following sum of SU(2) potentials
\begin{equation}
\label{G22}
e_{\mathrm{SU}(3)}[a] = \sli_{\sigma > 0} e_{\mathrm{SU}(2)(\vec{\sigma})}[a]  \, ,
\end{equation}
where the sum runs over all positive non-zero roots given in eq.~(\ref{G21}). 
SU(3) has two generators of the Cartan algebra ($t_3$, $t_8$). Therefore the effective 
potential of the Polyakov loop depends on two background fields $a_3$ and $a_8$ 
and is shown in fig.~\ref{fig13} for two temperatures, one below and one above the 
critical one.
As one observes in these figures both for $T < T_C$ and $T > T_C$ an absolute 
minimum occurs for $a_8 = 0$. Cutting the effective potential along the $a_8 = 0$ 
axis one finds the temperature dependent potential shown in fig.~\ref{fig12} (b),
which exhibits a first order phase transition. With a Gribov mass of 
$M = 880 \, \mathrm{MeV}$ as input one finds from the self-consistent (zero-temperature) 
solution a critical temperature of $T_C = 283 \, \mathrm{MeV}$.

Finally, if one includes fermions the deconfinement phase transition is turned into 
a crossover for both SU(2) and SU(3) \cite{34}. 

\section{Summary and Conclusions}

In my talk I have summarized basic features of the Hamiltonian approach to 
Yang--Mills theory in Coulomb gauge. I have shown how the
Gribov--Zwanziger confinement scenario is realized in this approach, and I 
have also established 
the connection with two alternative pictures of confinement, namely, the
condensation of magnetic monopoles (dual Mei{\ss}ner effect) and the center 
vortex picture. In particular, I have shown shown that the Coulomb string tension 
is not related to the temporal but rather to the spatial Wilsonian string tension 
and hence has to increase with the temperature above the deconfinement phase transition. 
I have then extended the 
Hamiltonian approach to finite temperatures in two ways, first by means of the 
standard grand canonical ensemble and second by compactifying one spatial 
dimension. The latter formulation is advantageous since it does not require any 
assumption for the density operator of the grand canonical ensemble; instead the 
finite temperature theory is fully encoded in the vacuum state calculated on the 
partially compactified spatial manifold \mbox{$\mathbb{R}^2 \times S^1(L)$}. 
Within this approach the effective potential of the Polyakov loop was calculated 
and a second and first, respectively, order phase transition was found for the 
gauge group SU(2) and SU(3). These phase transitions turn into a crossover when
the quarks are included. All these features are also corroborated by 
lattice calculations.

\section*{Acknowledgement}

This work was supported by DFG under contract no.~DFG-Re856/9-2.

\end{document}